\pdfoutput=1 
\documentclass[lettersize,journal]{IEEEtran}
\usepackage{amsmath,amsfonts}
\usepackage{algorithmic}
\usepackage{algorithm}
\usepackage{array}
\usepackage[caption=false,font=normalsize,labelfont=sf,textfont=sf]{subfig}
\usepackage{textcomp}
\usepackage{stfloats}
\usepackage{url}
\usepackage{verbatim}
\usepackage[numbers,sort&compress]{natbib}
\usepackage{graphicx}
\usepackage{amssymb}
\usepackage{cleveref}
\usepackage{pdfpages}
\def\BibTeX{{\rm B\kern-.05em{\sc i\kern-.025em b}\kern-.08em
    T\kern-.1667em\lower.7ex\hbox{E}\kern-.125emX}}
\usepackage{balance}
\usepackage[T1]{fontenc}

\begin{document}

\title{Low-complexity Beam Selection algorithms based on SVD for MmWave Massive MIMO Systems}
\author{Jinxing Yang, Jihong Yu, Shuai Wang, Hao Liu}



\maketitle

\begin{abstract}
To realize mmWave massive MIMO systems in practice, Beamspace MIMO with beam selection provides an attractive solution at a considerably reduced number of radio frequency (RF) chains. We propose low-complexity beam selection algorithms based on singular value decomposition (SVD). We first diagonalize the channel matrix by SVD, and the appropriate beams are selected one-by-one in a decremental or incremental order based on the criterion of sum-rate maximization. To reduce the complexity of the proposed algorithms significantly, we make use of SVD in the last iteration to aviod SVD from scratch again. Meanwhile, our proposed algorithms naturally obtain the precoding matrix, which can eliminate the multiusers interference. Simulation results demonstrate that our proposed algorithms can outperform the competing algorithms, including the fully digital zero-precoding.

\end{abstract}


\section{Introduction}

The rapid development of mobile networks and Internet of Things (IoT) technologies will generate a large amount of data. To meet the explosive capacity demand for future wireless communications, millimeter-wave (mmWave) communication is a promising technology which operates in the spectrum between 30 GHz and 300 GHz. However, since each antenna must be connected to one dedicated RF chain for mmWave MIMO, there is unbearable power consumption and hardware cost in massive multiple-input multiple-output (MIMO) scenarios relying on the numerous antennas\cite{brady2013beamspace}.

To reduce the required number of RF chains, traditional MIMO channel can be transformed into beamspace MIMO (B-MIMO) channel \cite{zeng2016millimeter} by employing a designed discrete lens array (DLA). As DLA plays the role of convex lens, the signals converge at different points of the focal surface, and it leads to the sparse channel with angle-dependent energy-focusing capabilities in B-MIMO \cite{amadori2015low}. Due to sparse structure of B-MIMO \cite{amadori2015low}, We can only select a few energy-focused beams with negligible performance losses \cite{sayeed2013beamspace}, and the required number of RF chains can be drastically fewer than traditional MIMO. However, it has been proved in \cite{liu2020statistical} that the beam selection problem is NP-hard.

To address this challenging problem, the maximum magnitude beam selection algorithm (referred as "MM-BS") proposed in \cite{sayeed2013beamspace} selects the beams with largest magnitude for each user which may lead to multiple users selecting the same beam. Furthermore, the interference aware beam selection algorithm in \cite{gao2016near} (referred as "IA-BS") circumvents this problem and achieves higher performance than MM-BS by considering the potential interference among users. Several algorithms based on different criterias have been proposed in \cite{amadori2015low}, such as maximizing signal-to-interference-ratio (SINR) and maximizing capacity. Based on QR decomposition of the beamspace channel matrix, a beam selection algorithm and associated precoding matrix have been proposed in \cite{pal2018beam} (referred as "QRD-BS"). Though it achieves superior performance than other algorithms, it suffers from very high computational complexity. To reduce the complexity of QRD-BS, a complexity-reduced beam selection algorithm for QRD-BS without degrading the system performance has been proposed in \cite{zhang2021complexity} (referred as "RQRD-BS"). 

Since RQRD-BS still exhibits high computational complexity, we propose novel SVD-based low-complexity beam selection algorithms. The low-complexity of the proposed algorithms is reflected in two aspects:

\begin{itemize}
\item[$\bullet$]
We first choose the fixed number of high energy beams to get a reduced-dimensional channel
\end{itemize}

\begin{itemize}
\item[$\bullet$]
To maximize sum-rate, the appropriate beams are selected one-by-one by SVD in a decremental or incremental order. To aviod SVD from scratch again, we achieve $\mathcal{O} (K^2)$ complexity for computing the sum-rate criterion. 
\end{itemize}

Moreover, the proposed algorithm can naturally get a more simple and effective precoding matrix than RQRD-BS without addition computation. The simulation shows that our proposed algorithms can outperform the aforementioned beam selection algorithms and the full dimensional zero-forcing precoding (referred as "FD-ZF").

\emph{Notations}: Matrices and vectors are denoted by boldface uppercase and lowercase letters, respectively. $a_{i}$ and $a_{i j}$ denote $i^{th}$ element of vector $\boldsymbol{a}$ and $(i j)^{th}$ element of matrix $\boldsymbol{A}$, respectively. $\boldsymbol {A}_{-j}$ denotes $\boldsymbol A$  with its $j^{th}$ row removed. $\boldsymbol I_M$ represents an $M \times M$ identity matrix. The superscripts $-1, T, H$ indicate inverse, transpose and conjugate transpose operator, respectively.

\section{SYSTEM MODEL}

A downlink mmWave MU-MIMO system is considered where the base station is equipped with $M$ transmit antennas and $N_{RF} \ll M$ RF chains to serve for $K$ single-antenna users. To ensure the spatial multiplexing gain, $N_{RF} \geq K$ should be satisfied. The precoded data vector $\boldsymbol x$ can be given by 
\begin{equation}
\boldsymbol{x}=\boldsymbol{P} \boldsymbol{s}=\sum_{k=1}^{K} \boldsymbol{p}_{k} s_{k}, \label{precoded_signal}
\end{equation}
where $\boldsymbol{s}= [s_1, \ldots, s_K]^T\in \mathbb{C}^{K\times 1}$ is the original signal vector with normalized power $\mathbb{E}\left(\boldsymbol{s s}^{H}\right)=\boldsymbol{I}_{K}$, $\boldsymbol{P}=\left[\boldsymbol{p}_{1}, \boldsymbol{p}_{2}, \ldots, \boldsymbol{p}_{K}\right] \in \mathbb{C}^{N_{RF} \times K}$ is the precoding matrix and $\boldsymbol{p}_{k}$ is the precoding vector for user $k$. 

The carefully designed DLA can be viewed as $M \times M$ discrete Fourier transform (DFT) matrix $\hat{\boldsymbol{U}}$, which can transforms the the conventional MIMO to the B-MIMO. Specifically, the DFT matrix $\hat{\boldsymbol{U}}$ includes array steering vectors with $M$ orthogonal directions spreading over the the entire space: 
\begin{equation}
\hat{\boldsymbol{U}}=\left[\boldsymbol{a}\left(\varphi_{1}\right), \boldsymbol{a}\left(\varphi_{2}\right), \cdots, \boldsymbol{a}\left(\varphi_{M}\right)\right]^{H}, \label{matrix_U}
\end{equation}
where $\boldsymbol{a}(\varphi)=\frac{1}{\sqrt{M}}\left[e^{-j 2 \pi \varphi i}\right]_{i \in \mathcal{I}(M)}$ denotes the $M \times 1$ array steering vector, and $\mathcal{I}(M)=\{i-(M-1) / 2, i=0,1, \cdots, M-1\}$. For DFT matrix $\hat{\boldsymbol{U}}$, we have $\varphi_{m}=\frac{1}{M}\left(m-\frac{M+1}{2}\right)$. Note that the matrix $\hat{\boldsymbol{U}}$ is orthonormal, i.e. $\hat{\boldsymbol{U}}^{H} \hat{\boldsymbol{U}}=\boldsymbol{I}$. Then received signal vector $\boldsymbol{y}$  of all $K$ users is expressed as
\begin{equation}
\boldsymbol{y}=\hat{\boldsymbol{H}}^{H} \boldsymbol{F P} \boldsymbol{s}+\boldsymbol{n}, \label{received_signal}
\end{equation}
where $\boldsymbol{F} \in$ $\mathbb{R}^{M \times N_{RF}}$ is the beam selection matrix whose entries $f_{i j}$ are either 0 or 1 , $\boldsymbol{n} \sim \mathcal{C} \mathcal{N}\left(\boldsymbol{0}, \sigma^{2} \boldsymbol{I}_{K}\right)$ is a $K \times 1$ additive white Gaussian noise (AWGN) vector and $\hat{\boldsymbol{H}} \in \mathbb{C}^{{M} \times K}$ is the beamspace channel matrix obtained by
\begin{equation}
    \hat{\boldsymbol{H}}=\left[\hat{\boldsymbol{h}}_{1}, \ldots, \hat{\boldsymbol{h}}_{K}\right]=[\hat{\boldsymbol{U}} \boldsymbol{g}_{1}, \hat{\boldsymbol{U}} \boldsymbol{g}_{2}, \ldots, \hat{\boldsymbol{U}} \boldsymbol{g}_{K}] \label{beamspace_channel},
\end{equation}
where $\{\boldsymbol{g}_{k}\}$ is narrowband clustered channel representation, based on the geometric channel model \cite{yu2019novel}, i.e.,
\begin{equation}
\boldsymbol{g}_{k}=\beta_{k}^{0} \boldsymbol{a}(\phi_{k}^{0})+\sqrt{\frac{1}{N_{cl}N_{ray}}}\sum_{i=1}^{N_{cl}} \sum_{l=1}^{N_{ray}} \beta_{k}^{il} \mathbf{a}\left(\phi_{k}^{il}\right) \label{SV_channel},
\end{equation}
where $N_{cl}$ be the number of scattering clusters and each cluster is composed of $N_{{ray}}$ subpaths. $\beta_{k}^{0}$ and $\beta_{k}^{il}$ represent the complex gains of the light of sight (LoS) and non-LoS (NLoS) components respectively. For the typical uniform linear array (ULA), the parameters $\phi_{k}^{0}$ and $\phi_{k}^{il}$ denote the spatial direction defined as $\phi=\frac{d}{\lambda} \sin \theta$, where $\theta$ is physical direction angle of propagation, $\lambda$ is the signal wavelength and $d$ is the antenna spacing, typically chosen as $d=\lambda / 2$. Thus the SINR of user $k$ can be expressed as
\begin{equation}
\gamma_{k}=\frac{\left|\hat{\boldsymbol{h}}_{k}^{H}  \boldsymbol{p}_{k}\right|^{2}}{\sum_{i \neq k}^{K}\left|\hat{\boldsymbol{h}}_{k}^{H}  \boldsymbol{p}_{i}\right|^{2}+\sigma^{2}}. \label{SNR}
\end{equation}
Then we can get sum-rate $R_{s}=\sum_{k=1}^{K} R_{k}$, where $R_{k}= \log_{2}(1+\gamma_{k})$ bits/s/Hz is the data rate achieved by user $k$. The problem can be formulated as
\begin{align}
\max_{\{\boldsymbol{F}, \boldsymbol{P}\}} \quad & R_s \label{joint_problem} 
\\ 
    \mbox{s.t.}\quad
 & \operatorname{Tr}\left(\boldsymbol{P}^{H} \boldsymbol{F}^{T} \boldsymbol{F} \boldsymbol{P}\right) \leq \rho  \tag{\ref{joint_problem}{a}}\label{joint_problema}\\
    &\sum_{i=1}^{M}\boldsymbol{F}_{i j}=1, \quad \forall j \tag{\ref{joint_problem}{b}} \label{joint_problemb}\\
    &\sum_{j=1}^{N_{RF}} \boldsymbol{F}_{i j} \leq 1, \quad \forall i \tag{\ref{joint_problem}{c}} \label{joint_problemc}
\end{align}
where $\rho$ is the transmit power budget. Note that each row of $\hat{\boldsymbol{H}}$ represents a beam vector. The constraints \eqref{joint_problemb} and \eqref{joint_problemc} mean that we need to select $N_{RF}$ beams from $M$ beams to serve $K$ users. In fact, it is a mixed-integer non-linear programming (MINLP) problem since it involves a discrete variable $F$ \cite{hu2021joint}.

\section{PROPOSED BEAM SELECTION ALGORITHM}
In this section,  We first outline \emph{motivation} behind the proposed algorithms which can help us understand the proposed algorithms in the following subsections. Next we introduce the proposed algorithms. 

Due to $K \ll M$, the beamspace channel matrix $\hat{\boldsymbol{H}} \in \mathbb{C}^{M \times K}$ is very high. In fact, we can select $N$ highest energy beams to form a new reduced-dimensional beamspace matrix $\boldsymbol H \in \mathbb{C}^{N \times K}$. Now we aim to select $N_{RF}$ beams out of $N$ beams rather than $M$ beams. Let the SVD decomposition \cite{kalman1996singularly} of $\boldsymbol{H}$ is given by 
\begin{equation}
    \boldsymbol{H} = \boldsymbol{U} \boldsymbol{\Sigma} \boldsymbol V^H,
\end{equation}
where $\boldsymbol{U}, \boldsymbol{V}\in \mathbb{C}^{N \times N}$ both are complex unitary matrixs and $\boldsymbol{\Sigma}$ is an $N\times K$ rectangular diagonal matrix with non-negative real numbers on the diagonal. The received signal vector $\hat{\boldsymbol{y}}$ without beam selection can be written as
\begin{equation}
\hat{\boldsymbol{y}}=\boldsymbol{V} \boldsymbol{\Sigma} \boldsymbol{U}^H  \hat{\boldsymbol{P}} \boldsymbol{s}+\boldsymbol{n}.\label{H_SVD}
\end{equation}
where $\hat{\boldsymbol P}$ is the precoding matrix. If we choose $\hat{\boldsymbol P} = \boldsymbol {U}, \tilde{\boldsymbol{y}} =\boldsymbol{V}^{H} \hat{\boldsymbol{y}}$ and $\tilde{\boldsymbol{n}}=\boldsymbol{V}^{H} \boldsymbol{n}$, we can rewrite the equation \eqref{H_SVD} as
\begin{equation}
    \tilde{\boldsymbol{y}}= \boldsymbol{\Sigma} \boldsymbol{s} +\tilde{\boldsymbol{n}}. \label{H_eq}
\end{equation}
From the equation \eqref{H_eq}, the interference is equal to zero for all users and we get an equivalent representation of $\boldsymbol H$ as a parallel Gaussian channel. Thus the sum-rate $R_s$ with uniform power allocation is given by
\begin{equation}
    R_{s}=\sum_{k=1}^K \log _{2}\left(1+\frac{1}{N_{0}} \frac{\rho}{K} \sigma_k^{2}\right) \text { bits } / \mathrm{s} / \mathrm{Hz} \label{sumate}
\end{equation}
where $\sigma_k$ is $(kk)^{th}$ element of $\boldsymbol{\Sigma}$ and $N_{0}$ is the noise variance of the AWGN. 

\subsection{Simplified SVD-based algorithm}
From equation \eqref{sumate}, the sum-rate $R_s$ can be estimated as
\begin{align}
R_{s}=\sum_{k=1}^K \log _{2}\left(1+\frac{1}{N_{0}} \frac{\rho}{K} \sigma_k^{2}\right) \leq \log_{2} \left(1+\frac{1}{N_{0}} \frac{\rho}{K} \sum_{k=1}^K \sigma_k^{2}\right)
\end{align}
Due to
\begin{equation}
    \sum_{k=1}^{K} \sigma_k^{2}=\operatorname{Tr}\left(\mathbf{\boldsymbol H^{H} \boldsymbol H}\right) \label{sigma_2} = \sum_{ij} |h_{ij}|^2, 
\end{equation}
simplified SVD-based algorithm (referred as "SSVD-BS") selects $K$ highest energy beams directly to maximize the upper bound of $R_s$, i.e.,  $\log_{2} \left(1+\frac{1}{N_{0}} \frac{\rho}{K} \sum_{k=1}^K \sigma_k^{2}\right)$. By this way, while SSVD-BS involves some performance loss to some extent compared to the other two algorithms proposed later, it enjoys the lowest computational complexity $\mathcal O{(NK)}$ since we only need to scan the channel matrix once.

\subsection{Decremental SVD-based algorithm}
We now propose a novel decremental SVD-based algorithm based on \eqref{sumate} (referrd as "DSVD-BS"). 
The DSVD-BS algorithm consists of $N-N_{RF}$ iterations. We denote $\boldsymbol{H}^{(i)}$ as the channel matrix at the begin of $i^{th}$ iteration. In iteration $i$, we need to eliminate a beam (i.e., a row of $\boldsymbol{H}^{(i)}$) with the minimum sum-rate loss, which can be explained as follows.

For $j=1, \ldots, N-i$, we remove the $j^{th}$ row of $\boldsymbol{H}^{(i)}$ to get the matrix $\boldsymbol{H}_{-j}^{(i)}$, compute its SVD decomposition $\boldsymbol{H}_{-j}^{(i)}=\boldsymbol{U}_{-j}^{(i)} \boldsymbol{\Sigma}_{-j}^{(i)} (\boldsymbol V^{(i)}_{-j})^H $ and then calculate the sum-rate 
\begin{equation}
    \alpha_{j}^{(i)}=\sum_{u \in\{1, \ldots, N-i\}} \log_{2}\left(1+\frac{1}{N_{0}} \frac{\rho}{i+1} \left(\sigma_{-j_{u u}}^{(i)}\right)^{2}\right), \label{alpha}
\end{equation}
where $\sigma_{-j_{u u}}^{(i)}$ denotes the $(uu)^{th}$ element of $\boldsymbol{\Sigma}_{-j}^{(i)}$. We can eliminate the beam whose contribution on the sum-rate is the least, which means we remove $\tilde{j}^{th}$ row from $\boldsymbol{H}^{(i)}$ to obtain $\boldsymbol{H}^{(i+1)}$, where $\tilde{j}=\arg \max _{j \in\{1, \ldots, N-i\}} \alpha_{j}^{(i)}$. The above Algorithm 1 depicts this process.

\begin{algorithm}[t]
	\renewcommand{\algorithmicrequire}{\textbf{Input:}}
	\renewcommand{\algorithmicensure}{\textbf{Output:}}
    \newcommand\sIf[2]{ \If{#1}#2\EndIf}          
	\caption{Decremental SVD-based algorithm}
	\label{alg2}
	\begin{algorithmic}[1]
		\REQUIRE $\boldsymbol H$
		\STATE Initialize $\boldsymbol H^{(0)} = \boldsymbol H$
		\FOR{$i=0, 1,\ldots, N-N_{RF}-1$}
		    \FOR{ $j=1, 2,\ldots, N-i$}
		        \STATE Remove $j^{th}$ row from $\boldsymbol{H}^{(i)}$ to get $\boldsymbol{H}_{-j}^{(i)}$
		        \STATE Compute $\boldsymbol{H}_{-j}^{(i)}=\boldsymbol{U}_{-j}^{(i)} \boldsymbol{\Sigma}_{-j}^{(i)} (\boldsymbol V^{(i)}_{-j})^H $
		        \STATE $\alpha_{j}^{(i)}= \sum_{u \in\{1, \ldots, N-i\}} \log_{2}\left(1+\frac{1}{N_{0}} \frac{\rho}{i+1} \left(\sigma_{-j_{u u}}^{(i)}\right)^{2}\right)$
		    \ENDFOR
		  \STATE $\tilde{j} =\arg \max _{j \in\{1, \ldots, N-i\}} \alpha_{j}^{(i)}$
		  \STATE Remove $\tilde{j}^{th}$ row from $\boldsymbol{H}^{(i)}$ to get $\boldsymbol{H}^{(i+1)}$ 
        \ENDFOR
    \ENSURE $\boldsymbol{H}^{(N-N_{RF})}$ 
	\end{algorithmic}  
\end{algorithm}

\subsection{Incremental SVD-based algorithm}
Incremental SVD-based algorithm (referred as "ISVD-BS") consists of $N_{RF}$ iterations. In iteration $i^{th}$ iteration, ISVD-BS selects the beams (i.e., a row of $\boldsymbol{H}^{(i)}$) by following iterative process, whose contribution in terms of sum-rate is the highest. 

Let $\boldsymbol H^{(i)}_s$ is the matrix formed by the beams that were previously selected at the end of $(i-1)^{th}$ iteration. For $j=1, \ldots, N-i$, we append the $j^{th}$ row of $\boldsymbol{H}^{(i)}$ to $\boldsymbol{H}^{(i)}_s$ which is denoted by $\boldsymbol{H}_{j}^{(i)}$, i.e., 
\begin{equation}
    \boldsymbol{H}_{j}^{(i)} = \begin{pmatrix}\boldsymbol{H}^{(i)}_s \\ (\boldsymbol{h}^{(i)}_j)^T \end{pmatrix}, \label{R}
\end{equation}
where $\boldsymbol{h}^{(i)}_j$ is $j^{th}$ row of $\boldsymbol{H}^{(i)}$. We can compute SVD decomposition $\boldsymbol{H}_{j}^{(i)}=\boldsymbol{U}_{j}^{(i)} \boldsymbol{\Sigma}_{j}^{(i)} (\boldsymbol V_{j}^{(i)})^H $ and then calculate the sum-rate 
\begin{equation}
    \eta_{j}^{(i)}=\sum_{u \in\{1, \ldots, N-i\}} \log_{2}\left(1+\frac{1}{N_{0}} \frac{\rho}{K} \left(\sigma_{j_{u u}}^{(i)}\right)^{2}\right), \label{eta}
\end{equation}
where $\sigma_{j_{u u}}^{(i)}$ denotes the $(uu)^{th}$ element of $\boldsymbol{\Sigma}_{j}^{(i)}$. From this process, we can select the beam whose effect on the sum-rate is the highest, i.e.,  
\begin{equation}
    \hat{j} =\arg \max _{j \in\{1, \ldots, N-i\}} \eta_{j}^{(i)}
\end{equation}
Then we can append the $\hat{j}^{th}$ row of $\boldsymbol{H}^{(i)}$ to $\boldsymbol{H}^{(i)}_s$ to get $\boldsymbol{H}_{s}^{(i+1)}$ and remove $\hat{j}^{th}$ row of $\boldsymbol{H}^{(i)}$ to get $\boldsymbol{H}^{(i+1)}$. The above Algorithm 2 describes this process.

\begin{algorithm}[t]
	\renewcommand{\algorithmicrequire}{\textbf{Input:}}
	\renewcommand{\algorithmicensure}{\textbf{Output:}}
    \newcommand\sIf[2]{ \If{#1}#2\EndIf}          
	\caption{Incremental SVD-based algorithm for beam selection}
	\label{alg3}
	\begin{algorithmic}[1]
		\REQUIRE $\boldsymbol H$
		\STATE Initialize $\boldsymbol H^{(0)} $ to $ \boldsymbol H$ and $\boldsymbol{H}_s^{(i)} $ to a empty matrix
		\FOR{$i=0, 1,\ldots, N_{RF}-1$}
		    \FOR{ $j=1, 2,\ldots, N-i$}
		        \STATE Obtain $\boldsymbol{H}_{j}^{(i)}$ based on \eqref{R}
		        \STATE Compute $\boldsymbol{H}_{j}^{(i)}=\boldsymbol{U}_{j}^{(i)} \boldsymbol{\Sigma}_{j}^{(i)} (\boldsymbol V^{(i)}_{j})^H $
		        \STATE $\eta_{j}^{(i)}= \sum_{u \in\{1, \ldots, N-i\}} \log_{2}\left(1+\frac{1}{N_{0}} \frac{\rho}{N-i} \left(\sigma_{j_{u u}}^{(i)}\right)^{2}\right)$
		    \ENDFOR
		  \STATE $\hat{j} =\arg \max _{j \in\{1, \ldots, N-i\}} \eta_{j}^{(i)}$
		  \STATE Append the $\hat{j}^{th}$ row of $\boldsymbol{H}^{(i)}$ to $\boldsymbol{H}_s^{(i)}$ to get $\boldsymbol{H}_{s}^{(i+1)}$ 
		  \STATE Remove $\hat{j}^{th}$ row of $\boldsymbol{H}^{(i)}$ to get $\boldsymbol{H}^{(i+1)}$.
        \ENDFOR
    \ENSURE $\boldsymbol{H}^{(N_{RF})}_s$ 
	\end{algorithmic}  
\end{algorithm}

\subsection{Complexity reduction}
For DSVD-BS algorithm, We first omit the superscript $(i)$ for readability. Without loss of generality, we assum that $N \geq 2*N_{RF}$. We denote the diagonal elements of $\boldsymbol{\Sigma}^2$ and $\boldsymbol{\Sigma}_{-j}^2$ as $\boldsymbol{D}=diag\{{d_{1}, \cdots, d_K}\}$ and $\boldsymbol{D}^{\prime}= diag\{{d_{1}^{\prime}, \cdots, d_K^{\prime}}\}$ respectively, where $d_{1} > \cdots > d_{K}$ and $d_{1}^{\prime} > \cdots > d_{K}^{\prime}$. To reduce the complexity of DSVD-BS, our goal is to make use of the prior knowledge $\{{d_{1}, \cdots, d_K}\}$  to aviod computing $\{{d_{1}^{\prime}, \cdots, d_K^{\prime}}\}$ from scratch again.

Based on $\boldsymbol{H}_{-j}^H \boldsymbol{H}_{-j} = \boldsymbol{H}^H \boldsymbol{H}-\boldsymbol{h}_j \boldsymbol{h}_j^{H}$ and eigenvalue decomposition (EVD) of $\boldsymbol{H}_{-j}^H \boldsymbol{H}_{-j}$ and $\boldsymbol{H}^H \boldsymbol{H}$, we can get
\begin{equation}
    \boldsymbol{V}_{-j} \boldsymbol{D}^{\prime} \boldsymbol{V}_{-j}^H = \boldsymbol{V} \boldsymbol{D} \boldsymbol{V}^H-\boldsymbol{h}_j \boldsymbol{h}_j^{H} \label{EVD}.
\end{equation}
Let $\boldsymbol z=\boldsymbol V^{H} \boldsymbol{h}_j$, we have
\begin{equation}
    \boldsymbol{h}_j \boldsymbol{h}_j^H = \boldsymbol{Vz} \boldsymbol{z}^H \boldsymbol{V}^H. \label{zz}
\end{equation}
Substituting \eqref{zz} into \eqref{EVD}, then
\begin{equation}
    \boldsymbol{V}_{-j} \boldsymbol{D}^{\prime} \boldsymbol{V}^H_{-j} = \boldsymbol{V}(\boldsymbol{D} - \boldsymbol z \boldsymbol z^H) \boldsymbol V^H.
\end{equation}
Since $\boldsymbol{D} - \boldsymbol z \boldsymbol z^H$ is Hermitian, it has the EVD $\boldsymbol{D} - \boldsymbol z \boldsymbol z^H = \boldsymbol{QSQ}^H$. The EVD is unique since the eigenvalues are distinct. Therefore, 
\begin{align}
    \boldsymbol{V}_{-j} &= \boldsymbol{VQ}\\
    \boldsymbol{S} &= \boldsymbol{D}^{\prime}
\end{align}
According to \cite{golub2013matrix}, The eigenvalues of $\boldsymbol{D} - \boldsymbol z \boldsymbol z^H$ are the roots $d^{\prime}=d_{1}^{\prime}, \cdots, d_{K}^{\prime}$ of the secular function
\begin{equation}
    f(d') = 1-\left(\frac{|z_{1}|^{2}}{d_{1}-d'}+\cdots+\frac{|z_{K}|^{2}}{d_{K}-d'}\right), \label{secular1}
\end{equation}
and the eigenvector $\boldsymbol q_{i}$ corresponding to each $d_{i}^{\prime}$ is
\begin{align}
    \boldsymbol q_{i} = \frac{\left(\boldsymbol{D}-d_{i}^{\prime} \boldsymbol{I}\right)^{-1} \boldsymbol{z}}{\left\|(\boldsymbol{D}-d_{i}^{\prime} \boldsymbol{I})^{-1} \boldsymbol{z}\right\|} \text {. } \label{vector}
\end{align}
The function $f(d')$ is a monotonically decreasing function in between its poles because
\begin{equation}
    f^{\prime}(d') = -\left(\frac{|z_{1}|^{2}}{(d_{1}-d')^2}+\cdots+\frac{|z_{K}|^{2}}{(d_{K}-d')^2}\right)<0.
\end{equation}
Thus, the eigenvalues $d_{i}^{\prime}$ satisfy the following interlacing property:
\begin{equation}
    d_{k+1} < d_{k}^{\prime} < d_{k}, \quad k=1,2, \cdots, K
\end{equation}
with $d_{K+1}=d_{K}-|z|^{2}$. 

For the ISVD-BS algorithm, we have $\boldsymbol{H}_{j}^H \boldsymbol{H}_{j} = \boldsymbol{H}^H \boldsymbol{H}+\boldsymbol{h}_j \boldsymbol{h}_j^{H}$. Similar to the above process, the eigenvalues $\tilde{d_{k}}$ of $\boldsymbol{H}_{j}^H \boldsymbol{H}_{j}$ are the roots of the secular equation 
\begin{equation}
    g(\tilde{d}) = 1+\left(\frac{|z_{1}|^{2}}{d_{1}-\tilde{d}}+\cdots+\frac{|z_{K}|^{2}}{d_{K}-\tilde{d}}\right), 
\end{equation}
and satisfy 
\begin{equation}
    d_{k} \leq \tilde{d_{k}} \leq d_{k-1}, \quad k=1,2, \cdots, K \label{secular2}
\end{equation}
where $d_{0}=d_{1}+|z|^{2}$. 

To find the roots of equations \eqref{secular1} and \eqref{secular2}, we can apply bisection or the numerical algorithms in \cite{melman1997numerical}, i.e., we can compute $\{d_{1}^{\prime}, \cdots, d_{K}^{\prime}\}$ and $\{\tilde{d}_{1}, \cdots, \tilde{d}_{K}\}$ based on $\{{d_{1}, \cdots, d_K}\}$ with $\mathcal{O}{(K^2)}$ computational complexity. Meanwhile, we can get the eigenvectors corresponding to eigenvalues by the equation \eqref{vector} with $\mathcal{O}{(K^2)}$ computational complexity.

\subsection{Computational Complexity Analysis}
The main complexity of the above two algorithms is due to the SVD decomposition in the step 5. Since we can get $\boldsymbol D^{\prime}$ and $\tilde{\boldsymbol D}$ based on $\boldsymbol{D}$ directly rather than performing SVD, the step 5 in the Algorithm 1 and 2 can be omitted and calculate the criterion sum-rate: 

\begin{itemize}
\item[$\bullet$]
For DSVD-BS algorithm, we can first compute the SVD decomposition of $\boldsymbol{H}^{(0)} = \boldsymbol{U}^{(0)} \boldsymbol{\Sigma}^{(0)} (\boldsymbol{V}^{(0)})^H$. Then for the outer iterations $i=0, 1,\ldots, N-N_{RF}-1$ of DSVD-BS algorithm, the computational complexity is   $\mathcal{O}\left((N-i)K^2\right)$ since the step 6 need $\mathcal{O}\left(K^2\right)$ computational complexity. Therefore, the total computation complexity of DSVD-BS algorithm is $\mathcal O\left(\sum_{i=0}^{N-N_{RF}-1}(N-i)K^2\right)$.
\end{itemize}

\begin{itemize}
\item[$\bullet$]
For the outer iterations $i=0, 1,\ldots, N_{RF}-1$ of ISVD-BS algorithm, the computational complexity is $\mathcal{O}\left((N-i)K^2\right)$ since the step 6 need $\mathcal{O}\left(K^2\right)$ computational complexity. Therefore, the total computation complexity of ISVD-BS algorithm is $\mathcal O\left(\sum_{i=0}^{N_{RF}-1}(N-i)K^2\right)$.
\end{itemize}

From the above discussion, when $N \ge 2N_{RF}$, the complexity of DSVD-BS is higher than ISVD-BS roughly and vice versa. The complexity of QRD-BS \cite{pal2018beam} and RQRD-BS \cite{zhang2021complexity} algorithms are $ O\left(\sum_{i=0}^{M-N_{RF}-1}(M-i)^2 K^{2}\right)$ and $\mathcal{O}\left(\sum_{i=0}^{M-N_{RF}-1}(M-i)K^2+K(M-i)^2\right)$ respectively, which are much larger than DSVD-BS in three aspects:

\begin{itemize}
\item[$\bullet$] $M$ is much larger than $N$, which makes $(M-i)$ is much larger than $(N-i)$ for both algorithms.
\end{itemize}

\begin{itemize}
\item[$\bullet$] Compared to QRD-BS, the degree of the term $(N-i)$ in complexity of DSVD-BS is $1$. However, The degree of the corresponding term in complexity of QRD-BS is $2$.
\end{itemize}

\begin{itemize}
\item[$\bullet$] Compared to RQRD-BS, ours doesn't have the term $\mathcal{O}\left(\sum_{i=0}^{M-N_{RF}-1}K(M-i)^2\right)$.
\end{itemize}

\section{Simulation results}
In this section, we evaluate the performance of the aforementioned algorithms by performing numerical simulations. 
For the channel model \eqref{SV_channel} of user $k$, we set $M=256, K=24, N_{cl}=2, N_{ray}=5, \alpha_{k}^{0} \sim \mathcal{C N}(0,1)$ and $\alpha_{k}^{il} \sim \mathcal{CN}\left(0, 10^{-1}\right)$. The SNR is set to $30$ dB. The $\theta_{k}^{0}$ and $\phi_{k}^{il}$ are sampled from i.i.d. uniform distribution within $[-0.5, 0.5]$. Since performance of ISVD-BS and DSVD-BS is almost the same when $N=3N_{RF}$, here we only show the performance of ISVD-BS. The simulation results are averaged over 1000 channel realizations.

Fig. 1 depicts the sum-rate performance versus the SNR of different beam selection algorithms, including FD-ZF, IA-BS, RQRD-BS, SSVD-BS and ISVD-BS algorithms. It's obvious that ISVD-BS outperforms other beam selection algorithms. SSVD-BS involves some performance loss to some extent compared to ISVD-BS and the performance gap becomes more pronounced as the SNR increases. It is worth noting that ISVD-BS can achieve better performance with a much lower complexity than RQRD-BS. While SSVD-BS has the lowest complexity $\mathcal{O} (NK)$, SSVD-BS even can achieve very close performance of QRD-BS and outperform the performance of FD-ZF when the SNR is low. Thus, in the stringent computational complexity scenario, SSVD-BS becomes a very attractive algorithm for beam selection. 

\begin{figure}[htbp]
	\centering
    \includegraphics[page=1, width=0.9\columnwidth,height=0.5\columnwidth]{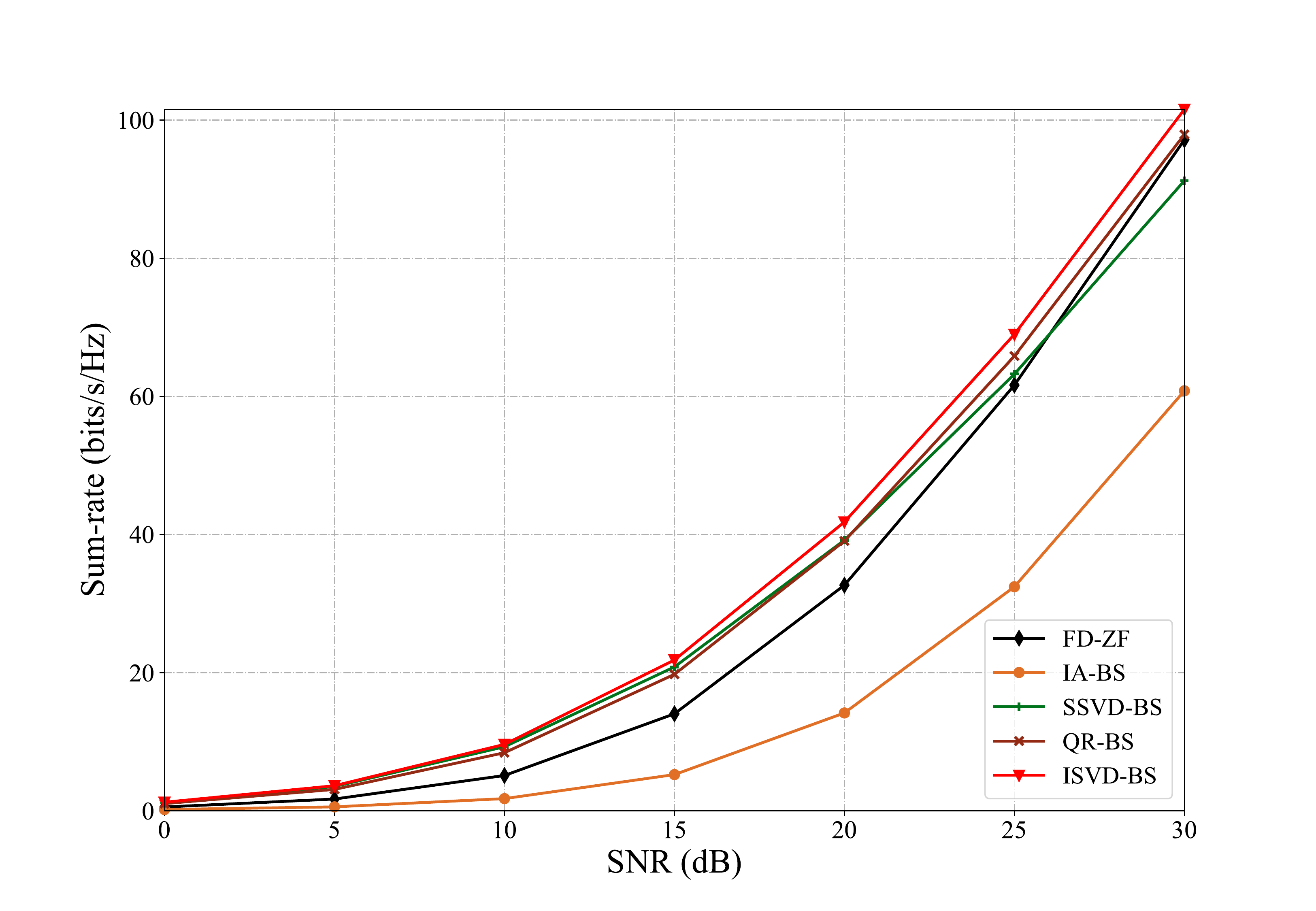}%
	\caption{Sum-rate performance verus the SNR.}
	\label{fig1}
\end{figure}

Fig. 2 shows the sum-rate comparison of different beam selection algorithms against the number of users $K$. We can see that for all five algorithms,  the sum-rate performance increases as the number of users increases, and our proposed ISVD-BS alogorithm can achieve the best performance, which shows its superiority to mitigate the multi-user interference. Besides, the performance gap between the proposed ISVD-BS alogorithm and the other algorithms as $K$ increases, which demonstrates its potentiality for applications with a large $K$. 

\begin{figure}[htbp]
	\centering
    \includegraphics[page=1, width=0.9\columnwidth,height=0.5\columnwidth]{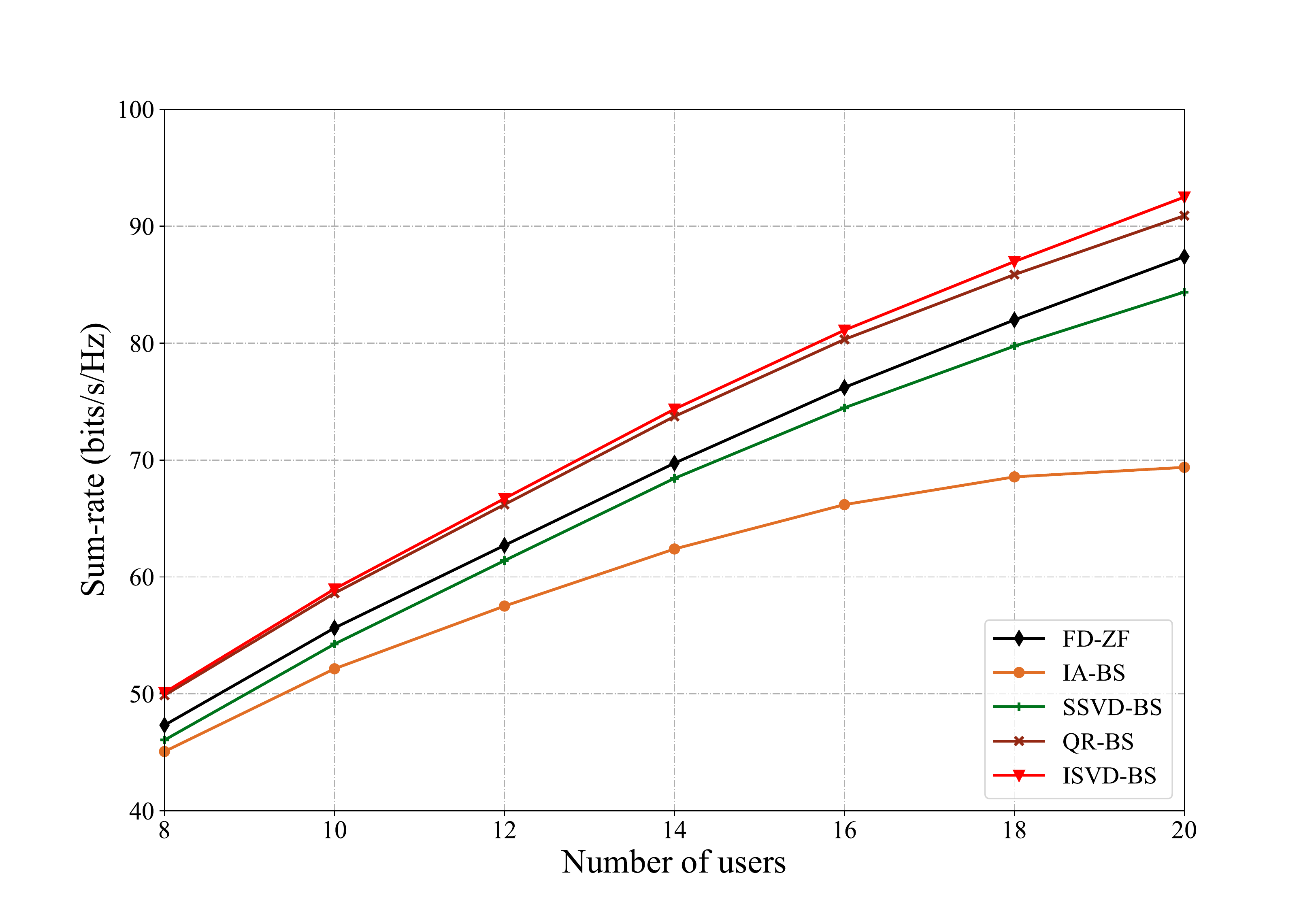}%
	\caption{Sum-rate performance verus the number of users $K$.}
	\label{fig2}
\end{figure}
In Fig. 3, we plot the sum-rate performance versus number of transmit antennas $M$, where $N_{RF}$ and $K$ are set to 16. ISVD-BS and SSVD-BS algorithms achieve much better performance compared to the other algorithms when $M = 20$. Meanwhile, the sum-rate of ISVD-BS improves solwly as $M$ increases, while the performance of IA-BS and FD-ZF are heavily dependent on $M$. Therefore, ISVD-BS can be deployed in the scenario where the number of antennas is limited, but its performance is still satisfactory enough. 

\begin{figure}[htbp]
	\centering
    \includegraphics[page=1, width=0.9\columnwidth,height=0.5\columnwidth]{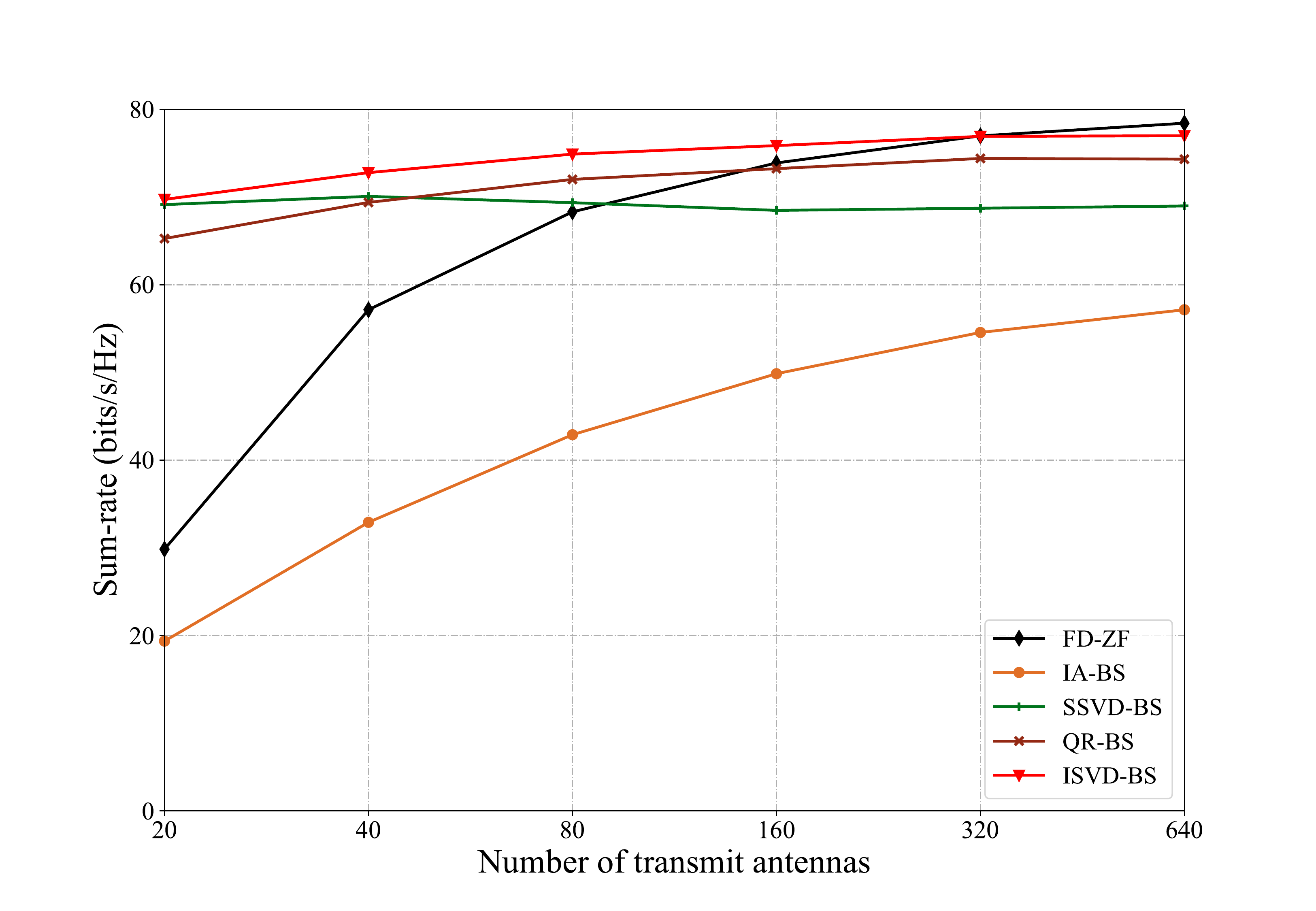}%
	\caption{Sum-rate performance verus the number of transmit antennas $M$.}
	\label{fig3}
\end{figure}

\begin{figure}[htbp]
	\centering
    \includegraphics[page=1, width=0.9\columnwidth,height=0.5\columnwidth]{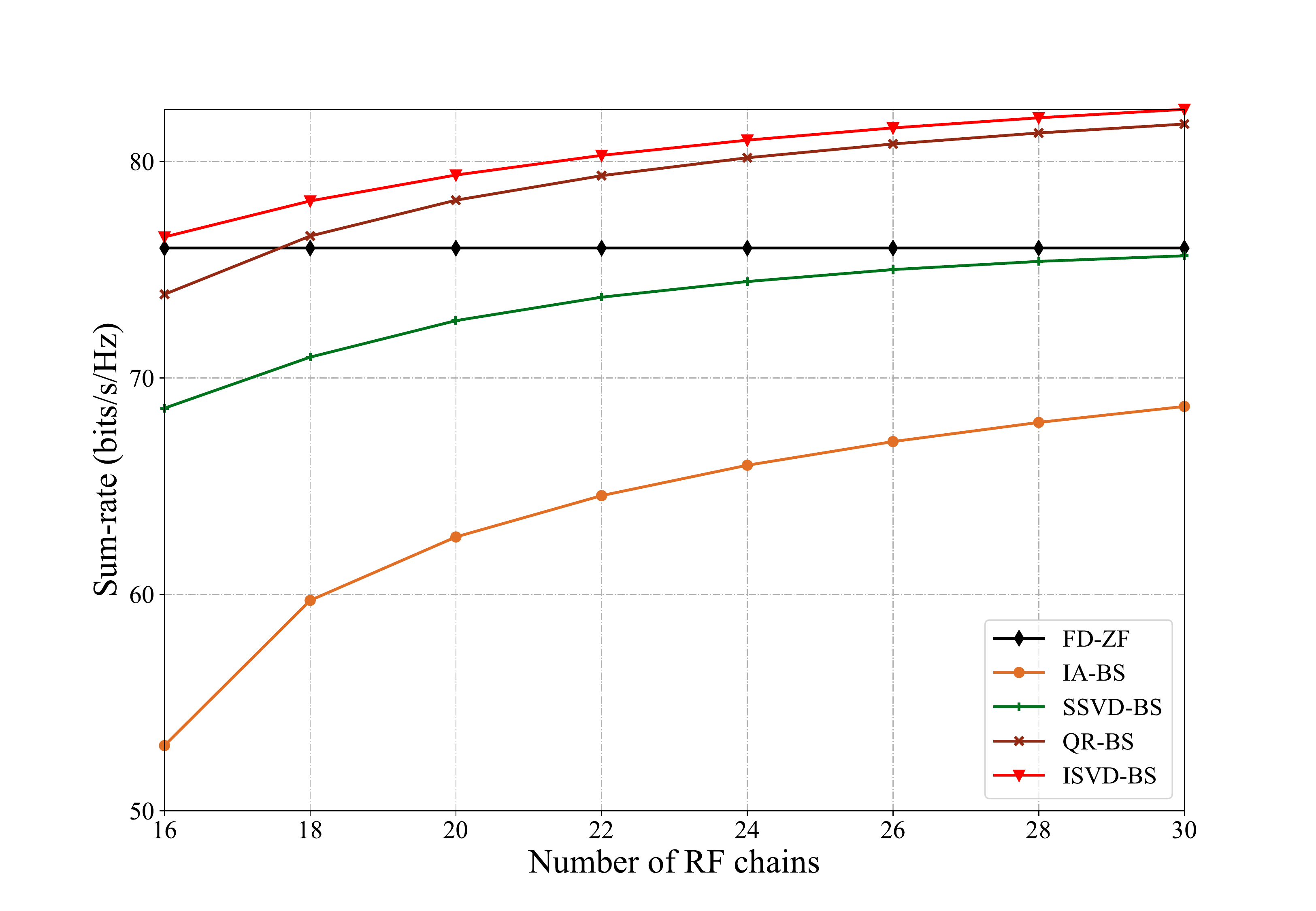}%
	\caption{Sum-rate performance verus number of RF $N_{RF}$.}
	\label{fig4}
\end{figure}

Fig. 4 illustrates the achievable system sum-rate versus
the number of RF $N_{RF}$ in different algorithms where $K=16$. We can see that the sum-rate achieved by all algorithms increases monotonically with $N_{RF}$.
Moreover, the ISVD-BS algorithm outperforms the QRD-BS algorithm when $N_{RF}$ is small, and the gap decreases with $N_{RF}$.
Since the use of each RF chain requires a certain amount of energy consumption, the smaller $N_{RF}$ means lower energy consumption. Therefore our ISVD-BS algorithm can achieve better performance significantly in the energy-constrained scenario than QR-BS.

\section{CONCLUSION}
In this letter, we proposed three low-complexity beam selection algorithms, i.e, SSVD-BS, DSVD-BS, and ISVD-BS  algorithms. Three algorithms all can achieve the better sum-rate performance than fully digital system with much higher energy efficiency. Particularly, the SSVD-BS almost has the lowest computation complexity $\mathcal{O}(NK)$. Meanwhile, DSVD-BS and ISVD-BS can outperform QRD-BS and RQRD-BS  with much lower computation complexity.
\balance

\small
\bibliographystyle{IEEEtran}
\bibliography{main_conciseV2.bbl}

\begin{thebibliography}{10}
\providecommand{\url}[1]{#1}
\csname url@samestyle\endcsname
\providecommand{\newblock}{\relax}
\providecommand{\bibinfo}[2]{#2}
\providecommand{\BIBentrySTDinterwordspacing}{\spaceskip=0pt\relax}
\providecommand{\BIBentryALTinterwordstretchfactor}{4}
\providecommand{\BIBentryALTinterwordspacing}{\spaceskip=\fontdimen2\font plus
\BIBentryALTinterwordstretchfactor\fontdimen3\font minus
  \fontdimen4\font\relax}
\providecommand{\BIBforeignlanguage}[2]{{%
\expandafter\ifx\csname l@#1\endcsname\relax
\typeout{** WARNING: IEEEtran.bst: No hyphenation pattern has been}%
\typeout{** loaded for the language `#1'. Using the pattern for}%
\typeout{** the default language instead.}%
\else
\language=\csname l@#1\endcsname
\fi
#2}}
\providecommand{\BIBdecl}{\relax}
\BIBdecl

\bibitem{brady2013beamspace}
J.~Brady, N.~Behdad, and A.~M. Sayeed, ``Beamspace mimo for millimeter-wave
  communications: System architecture, modeling, analysis, and measurements,''
  \emph{IEEE Transactions on Antennas and Propagation}, vol.~61, no.~7, pp.
  3814--3827, 2013.

\bibitem{zeng2016millimeter}
Y.~Zeng and R.~Zhang, ``Millimeter wave mimo with lens antenna array: A new
  path division multiplexing paradigm,'' \emph{IEEE Transactions on
  Communications}, vol.~64, no.~4, pp. 1557--1571, 2016.

\bibitem{amadori2015low}
P.~V. Amadori and C.~Masouros, ``Low {RF}-complexity millimeter-wave
  beamspace-{MIMO} systems by beam selection,'' \emph{IEEE Transactions on
  Communications}, vol.~63, no.~6, pp. 2212--2223, 2015.

\bibitem{sayeed2013beamspace}
A.~Sayeed and J.~Brady, ``Beamspace mimo for high-dimensional multiuser
  communication at millimeter-wave frequencies,'' in \emph{2013 IEEE global
  communications conference (GLOBECOM)}.\hskip 1em plus 0.5em minus 0.4em\relax
  IEEE, 2013, pp. 3679--3684.

\bibitem{liu2020statistical}
H.~Liu, X.~Yuan, and Y.~J. Zhang, ``Statistical beamforming for fdd downlink
  massive mimo via spatial information extraction and beam selection,''
  \emph{IEEE Transactions on Wireless Communications}, vol.~19, no.~7, pp.
  4617--4631, 2020.

\bibitem{gao2016near}
X.~Gao, L.~Dai, Z.~Chen, Z.~Wang, and Z.~Zhang, ``Near-optimal beam selection
  for beamspace mmwave massive mimo systems,'' \emph{IEEE Communications
  Letters}, vol.~20, no.~5, pp. 1054--1057, 2016.

\bibitem{pal2018beam}
R.~Pal, K.~Srinivas, and A.~K. Chaitanya, ``A beam selection algorithm for
  millimeter-wave multi-user mimo systems,'' \emph{IEEE Communications
  Letters}, vol.~22, no.~4, pp. 852--855, 2018.

\bibitem{zhang2021complexity}
Q.~Zhang, X.~Li, B.-Y. Wu, L.~Cheng, and Y.~Gao, ``On the complexity reduction
  of beam selection algorithms for beamspace mimo systems,'' \emph{IEEE
  Wireless Communications Letters}, vol.~10, no.~7, pp. 1439--1443, 2021.

\bibitem{yu2019novel}
H.~Yu, W.~Qu, Y.~Fu, C.~Jiang, and Y.~Zhao, ``A novel two-stage beam selection
  algorithm in mmwave hybrid beamforming system,'' \emph{IEEE Communications
  Letters}, vol.~23, no.~6, pp. 1089--1092, 2019.

\bibitem{hu2021joint}
Q.~Hu, Y.~Liu, Y.~Cai, G.~Yu, and Z.~Ding, ``Joint deep reinforcement learning
  and unfolding: Beam selection and precoding for mmwave multiuser mimo with
  lens arrays,'' \emph{IEEE Journal on Selected Areas in Communications},
  vol.~39, no.~8, pp. 2289--2304, 2021.

\bibitem{kalman1996singularly}
D.~Kalman, ``A singularly valuable decomposition: the svd of a matrix,''
  \emph{The college mathematics journal}, vol.~27, no.~1, pp. 2--23, 1996.

\bibitem{golub2013matrix}
G.~H. Golub and C.~F. Van~Loan, \emph{Matrix computations}.\hskip 1em plus
  0.5em minus 0.4em\relax JHU press, 2013.

\bibitem{melman1997numerical}
A.~Melman, ``A numerical comparison of methods for solving secular equations,''
  \emph{Journal of computational and applied mathematics}, vol.~86, no.~1, pp.
  237--249, 1997.

\end{thebibliography}

\end{document}